# Measurements of Polarization Dependencies in Parametric Down-Conversion of X-rays into Ultraviolet Radiation

S. Sofer[1], O. Sefi[1], A. G. A. Nisbet[2], and S. Shwartz[1]

[1]*Physics Department and Institute of Nanotechnology, Bar-Ilan University, Ramat Gan, 52900*

*Israel*

[2] *Diamond Light Source, Harwell Science and Innovation Campus, Didcot OX11 0DE, United*

*Kingdom*

We present measurements of the polarization dependencies of the x-ray signal photons generated by the effect of parametric down-conversion of x rays into ultraviolet radiation. The results exhibit pronounced discrepancies with the classical model for the nonlinearity but qualitatively agree with a newly developed quantum mechanical theory for the nonlinear interaction. Our work shows that the reconstruction of the atomic scale charge distribution of valence electrons in crystals by using nonlinear interaction between x rays and longer wavelength radiation, as was suggested in previous works, requires the knowledge of polarization of the generated x-ray signal beam. The results presented in this work indicate a new methodology for the study of properties of the Wannier functions in crystals.

The advent of high brilliance x-ray sources, such as third-generation synchrotron radiation facilities and x-ray free electron lasers (XFELs) has made it possible to study nonlinear interactions between x-rays and optical or ultraviolet (UV) radiation. One of the more intriguing opportunities in this field is studying the properties of valence electrons with atomic scale resolution, as was suggested theoretically, almost fifty years ago [1-3]. In recent years, several of those mixing processes have been observed. Sum-frequency generation (SFG) has been observed using an XFEL source [4], and the effect of difference frequency generation (DFG) has been studied theoretically [5-6]. The most studied nonlinear process, however, is the process of parametric down-conversion (PDC) of x rays into optical or UV radiation [7-15].

PDC of x rays into optical or UV radiation is a second order nonlinear effect where an input x-ray beam (denoted as pump) interacts with the vacuum fluctuations in a nonlinear medium to generate photons pairs, an x-ray photon, which is denoted as signal, and a photon at longer wavelengths, in the optical or UV regime, which is denoted as the idler. In this process both energy and momentum are conserved. Energy conservation dictates that the sum of the angular frequencies of the signal and idler is equal to the angular frequency of the pump ($\omega_p = \omega_s + \omega_{id}$, where $\omega_p, \omega_s, \omega_{id}$ are the angular frequencies of the pump, signal and idler, respectively). Since the distance between the atomic planes is comparable to the wavelengths of the x-ray fields, we utilize the reciprocal lattice vector for phase matching, which takes the form of $\vec{k}_p + \vec{G} = \vec{k}_s + \vec{k}_{id}$, where $\vec{k}_p, \vec{k}_s, \vec{k}_{id}$ are the wave vectors of the pump, signal and idler, respectively, and $\vec{G}$ is the reciprocal lattice vector.

Recent observation of PDC of x rays into longer wavelengths in non-centrosymmetric crystals indicated a strong nonlinear interaction, which was not predicted by early theories [1-3,14]. Those results indicated a new source for the nonlinearity, which was not taken into account in those theoretical works. More recently more comprehensive theories for the nonlinear interaction have been presented [16-18] and can be used to study the new results.

One of the recent theoretical works studied the nature of the nonlinear interaction between x rays and longer wavelength fields, while treating the nonlinear medium quantum mechanically, and taking into account the effect of the band-structure and the joint density of states on the nonlinear process [16]. The theory expresses the nonlinear conductivity in the wave mixing process as the summation over matrix elements in the Wannier basis of the electrons in the crystal. One of the major predictions of the theory, is that the nonlinear current at the signal frequency is not necessarily parallel to the polarization of the incoming pump, and thus the signal photons are not necessarily polarized at the scattering plane, as is predicted by the early theories [16]. Consequently, the expression for the nonlinear conductivity can be separated into two terms: one term, which contributes when the polarizations of the signal and pump photons are parallel and another term, which contributes when the polarizations are perpendicular. Moreover, the first term contains the information about the Fourier components of the induced charge density in the crystal. Thus, the full reconstruction of the induced charge density, as was proposed in previous works [2,4,10], must be accompanied by a polarization rotation measurement of the signal photons.

Here, we report on the experimental measurement of the polarization dependence of the signal photons generated by the effect of PDC of x rays to UV radiation. We measure the count rates of the x-ray signal generated by PDC in a gallium arsenide (GaAs) crystal, while using a polarizer to measure the polarization dependence of the signal photons. We compare the polarization dependence of the signal photons to the polarization dependence of the Bragg reflected beam and observe a distinct shift in the polarization angle. The generalization of our work can lead to the development of a novel method for constructing the induced charge density in crystals with atomic scale resolution. Moreover, further understanding of the quantum mechanical model could lead to a method for studying Wannier functions and their symmetries.

We conducted the experiment described in this article on beamline I16 of the Diamond Light Source [19]. The schematic of the experimental setup is shown in Fig.1. In the experiment, we use a monochromatic and collimated synchrotron beam, which is polarized in the scattering (horizontal) plane. The beam illuminates a GaAs crystal, which is mounted on a goniometer. We implement a polarizer for the x-ray signal by using a multi-bounce channel cut silicon crystal, which is mounted at an incidence angle of approximately 45 degrees. Since the cross section of Bragg scattering is proportional to the Thompson cross section, for an incidence angle of 45 degrees, the portion of the reflected beam, which is polarized in the scattering plane of the silicon crystal, is highly suppressed. In order to achieve a Bragg angle of 45 degrees for the silicon crystal we use an incident beam with an energy of 8.388 keV and the Si(333) atomic planes. We change the orientation of the polarizer simply by rotating the silicon crystal in the direction indicated by the arrow in Fig. 1, thus changing its scattering plane and selecting different polarizations. The silicon crystal has a second important role; it operates as an analyzer to select the signal photons from the residual elastic beam that is scattered from the GaAs crystal. This function is achieved since the polarizer is a crystal that acts as a Bragg filter. The overall energy resolution of the system is about 1 eV (and mainly determined by the energy bandwidth of the input monochromator). We note that since the rocking curve of the silicon analyzer crystal is very narrow, it is important to verify the correct position of the analyzer after each variation of the scattering plane. The silicon analyzer is thus calibrated after each rotation with respect to the Bragg elastic beam. After the analyzer crystal the photons are collected by a 2D pixelated detector (MerlinEM).

We begin by examining the recorded intensity distribution on the 2D detector in order to verify that the measured signal cannot be attributed to other known effects. Fig. 2 shows the angular and spectral distribution of the signals on the 2D detector. The vertical axis of the detector is converted to energy deviation from the pump energy and the horizontal axis is converted to the angle relative to the Bragg angle of the elastic signal (at the photon energy of the input beam). In the figure, each pixel corresponds to approximately 0.0025 degrees. The conversion is done by considering the pixel size (55μm) and the distance between the crystal and the detector (1.25 m). There are several potential scattering effects that can be observe in an experiment like we perform. The first is the residual Bragg scattering of the input beam that is not completely suppressed by the analyzer. Another potential effect is the so called "truncation rods", which are result of Bragg scattering from the surface of the crystal [20]. A clear distinction between these effects and the PDC is angular distribution shown on the 2D detector at a given input angle. The width of the elastically scattered beams is determined by the footprint of the input beam on the crystal convolved with its divergence. In our experiment the width of the input beam is approximately 100 μm and the beam divergence is about 11 mrad [19]. Thus, the elastically scattered beam spans about 10 pixels on the 2D detector, which translates into about 0.5 mrad as in images shown in Fig. 2. In contrast, the angular spread of the PDC signal is vastly broader since it is generated from a large number of vacuum fluctuation modes, which contribute to the measured signal. The photon energy width of the intrinsic PDC is also broad but is restricted on the detector by the width of the analyzer crystal. Hence, the shape of the PDC signal is approximately a cigar shape with the long dimension along the scattering plane. Figures 2a and 2b show the recorded patterns for the first and second solutions of the phase matching condition. Figures 2c and 2d, show the recorded data for the case where the detector is at the Bragg angle and for an angle where a truncation rod is visible. It is clear that the recorded patterns we interpret as PDC are much broader than any pattern expected from elastic scattering. The width of the PDC pattern is approximately 0.07 degrees (FWHM), while the widths of the Bragg and truncation rod patterns are approximately 0.005 degrees (FWHM). This result strengthens the claim that we indeed measure PDC and not an artifact originated from elastic scattering. Of importance, we note that in previous setups, where we used an analyzer with a lower energy resolution (Si(111)), the width of analog measurements was wider by up to a factor of 3, due to the one-to-one relation between the photon energy and the angle of propagation of the generated PDC photons, thus a broader energy acceptance leads to a broader angular width.

Next, we plot the efficiency (the sum over a region of interest on the 2D detector) of the measured signal photons as a function of the deviation of the detector from the Bragg angle and show that the observed peak positions agree with the phase matching condition of the PDC. In Fig. 3 we show the measured efficiency for an idler energy of 10 eV and with the pump angle deviated by 0.08 degrees with respect to the Bragg angle. Since the phase matching equations have two solutions, we expect to observe two peaks in the experiment, as has been reported in previous works [13] and is clearly seen in Fig. 3. In order to verify that the peak position is reasonable with the phase matching condition, we calculate the phase mismatch, according to $\Delta k_z L = (k_p \cos\theta_p - k_s \cos\theta_s - k_i \cos\theta_i)L$, where L is the absorption length at the idler wavelength and is taken to be 100 nm. The calculated phase mismatch is 2.27 rad for the first peak, and 0.375 rad for the second peak. Both values for the mismatch are much smaller than $2\pi$, and thus, are within the range for phase matching. Moreover, the two peaks are clearly separated from the Bragg peak, and cannot be explained by any other inelastic mechanism. We note that the exact angle for the phase matching is difficult to predict, due to the uncertainty in the refractive indexes in the UV regime, and the uncertainty introduced by the analyzer spectral width, which can correspond to an uncertainty of up to 10 mrad. Consequently, we conclude that we indeed measured the PDC signal. Similar values for the phase mismatch were found for the (2 0 0) and (3 3 3) atomic planes of the GaAs crystal.

After establishing that the signal we measure is generated by PDC, we measure the spectral dependence of the PDC process, as was done in a previous work [15] in order to verify that the observed features in the spectral dependence that can be attributed to the valence electrons, as expected from an effect that involves optical and UV waves. Figure 4 shows the spectral dependence of the PDC process for the (1 1 1) atomic planes. We observe two features, one near the bandgap at 1.4 eV and the second at at 6 eV, which is a transition energy in the band structure. The features in the spectral dependence, along with the general trend, are in good agreement with previous measurements of the spectral dependence for the same atomic planes in GaAs.

We now turn to measure the polarization dependence of the signal photons. To measure these dependencies, we start by finding the PDC signal, as was described previously. Then, we repeat this process for several angles of the polarization analyzer. We then plot the peak count rate of the PDC signal as a function of the angle of the polarizer. In order to measure the deviation of the polarization vector of the signal beam from the polarization vector of the pump beam, we compare the polarization dependence of the signal photons with the polarization dependence of the elastic beam. In Fig. 5, the polarization dependence of the signal photons is shown, for several atomic planes of the GaAs crystal. It is clearly visible that the polarization of the signal photons is different from the polarization of the elastically scattered beam. To quantify the deviation of the polarization vector of the signal photons from the polarization vector of the Bragg reflected beam, we fit the measured data for the PDC signal to a function of the form $\cos^2(\theta + \theta_0)$, where $\theta$ is the angle of the polarizer and $\theta_0$ is a fitting parameter, which determines the angular deviation from the Bragg reflected beam. We measure a deviation of 7.4 degrees for the (1 1 1) atomic planes, a deviation of 5.1 degrees for the (2 0 0) atomic planes, and a deviation of -17.6 degrees for the (3 3 3) atomic planes.

To understand the origin of the polarization dependence we observed, we consider the quantum mechanical model for the nonlinearity, which was described in Ref. [16]. The nonlinear conductivity of x rays and longer wavelength radiation can be expressed as:

$$\sigma_{ijk} = A_k(\omega_{id}, \vec{G}) \cdot \delta_{ij} + B_{ijk}(\omega_{id}, \vec{G}) \cdot (1 - \delta_{ij}), \tag{1}$$

where $\delta_{ij}$ is the Kronecker delta function, and:

$$A_k(\omega_{id}, \vec{G}) = \frac{\hbar e^3}{m^2 \omega_p V} \sum_{n_1, n_2} \langle W_{n_2} | e^{-i\vec{G}\cdot\vec{x}} | W_{n_1} \rangle \langle W_{n_1} | \vec{p} \cdot \hat{e}_k | W_{n_2} \rangle I_{n_2, n_1}(\omega_{id}, \vec{k}_{id}) \tag{2}$$

$$B_{ijk}(\omega_{id}, G) \tag{3}$$
$$= \frac{\hbar e^3 (\vec{k}_{id} - \vec{G})}{V m^3 \omega_p \omega_s} \sum_{n_1, n_2} \langle W_{n_2} | e^{-i\vec{G}\cdot\vec{x}} \cdot [\hat{e}_i (\vec{p}\cdot\hat{e}_j) - \hat{e}_j (\vec{p}\cdot\hat{e}_j)] | W_{n_1} \rangle \langle W_{n_1} | \vec{p}\cdot\hat{e}_k | W_{n_2} \rangle I_{n_1, n_2}(\omega_{id}, \vec{k}_{id}),$$

where $V$ is the volume of the crystal, $|W_n\rangle$ is the Wannier function of band $n$, $\vec{p}$ is the momentum operator, and $i, j, k$ are cartesian coordinates, where the nonlinear current is parallel to $\hat{e}_i$, $\hbar$ is the reduced Planck constant, $e$ is the electron charge, and $m$ is the electron mass. $I_{n_1, n_2}(\omega_{id}, \vec{k}_{id})$ is the spectral dependence of the interaction and is given by

$$I_{n_1, n_2}(\omega_{id}, \vec{k}_{id}) = \frac{V}{(2\pi)^3} \int_{B.Z.} d\vec{q} \frac{-\left(f_0\left(\varepsilon_{n_1}(\vec{q}+\vec{k}_{id})\right) - f_0\left(\varepsilon_{n_2}(\vec{q})\right)\right)}{\varepsilon_{id} \left[\left(\varepsilon_{n_1}(\vec{q}+\vec{k}_{id}) - \varepsilon_{n_2}(\vec{q}) - \varepsilon_{id}\right) + i\hbar\gamma_{n_1 \vec{q}+\vec{k}_{id}, n_2 \vec{q}}\right]}, \tag{4}$$

where $f_0(\varepsilon)$ is the Fermi-Dirac distribution, $\gamma_{nm}$ is the phenomenological damping coefficient, $\varepsilon_{n_1}(\vec{q})$ is the is the dispersion relation for electrons in band $n_1$, and the integration is over the Brillouin zone.

In Eq. 1, the first term contributes to the nonlinear current, which is parallel to the polarization of the pump beam, and the second term contributes to the nonlinear current which is perpendicular to the pump beam. The first term in the nonlinear conductivity is proportional to the Fourier component of the induced charge density. Both terms include information on the intermolecular interactions in the crystal (the Wannier matrix elements) and information on the band structure of the crystal, which is encoded in the spectral dependence function [16]. In contrast to the theory that has been used in Ref. [4,10] that predicts that the intensity of the signal beam is proportional to the induced charge of the valence electron, only the first term in Eq. 1 is attributed to the induced charge density. Another important difference is the prediction of the theory in Ref. [4,10] that polarization of the signal beam is always parallel to the polarization of the pump, while Eq. 1 contains two terms with different polarization dependencies.

According to Eq.1, the polarization dependence of the first term (which is proportional to the induced charge) is parallel to the polarization of the pump beam while the second term is normal to the polarization of pump beam. The implication of Eq. 1 is that the induced charge density of the valence electrons can be retrieved by polarization measurement. It is important to note that the second term, which is normal to the pump beam, is not necessarily normal to the scattering plane. However, with our polarizer we measure the polarization of the signal beam with respect to the scattering plane. Thus, we cannot deduce of the ratio between the two terms directly from our measurement and, but it requires a full fitting to the model and the parameters of the nonlinear medium. We note that the model predicts that the ratio between the part that is related to the induced charge to the part that is not related to the induced charge decreases as the angle between signal beam and the input beam approaches 90 degrees. This explains why good agreement between experiment and the classical theory has been observed in diamond; the angles between the signal and pump beam in those experiments were relatively small thus the induced charge term dominated. The polarization of the output beam was not measured in those experiments. Furthermore, since the solution of the phase matching equation is near the Bragg angle, the angle between the signal and the pump beams can be approximated to $2\theta_B$. It is indeed clearly visible from Fig. 5 that for a higher reflection, such as the GaAs(3 3 3) atomic planes, where the Bragg angle is about 40 degrees, the shift from the polarization of the Bragg beam is more prominent. Moreover, the information on the polarization of the PDC effect can be used to study the symmetries on the Wannier functions in the crystal. As can be seen in Eq. 2 and Eq. 3, the different contributions to the PDC effect, that can be separated by a polarization measurement, correspond to different matrix elements. This information can be utilized to study different types in intermolecular interactions, and even reconstruct the Wannier functions with atomic scale resolution. We note that a comprehensive reconstruction of the density of the valence electrons requires the measurement of a sufficient number of Fourier components of the nonlinear conductivity, and a full fit to the quantum mechanical theory, which requires complex simulations.

In conclusion, we have demonstrated the experimental observation of the polarization dependence of the signal photons generated by the effect of PDC of x rays to UV radiation in GaAs. The results we have measured suggest that the polarization dependence is not trivial, different from the polarization dependence of the elastic Bragg scattering, and cannot be explained by theories previously used for the description of the nonlinear interaction. A new quantum mechanical model for the nonlinearity predicts the polarization dependencies we observed [15]. An important conclusion from our work is that the polarization measurements for several Fourier components of the nonlinear conductivity is essential for the comprehensive study of atomic-scale induced charge density. Knowledge of the polarization dependence can be extremely important for measurement of the atomic-scale induced charge density in more exotic crystals that exhibit more complex phenomena, such as phase transitions, and charge density waves, since it can be more complex in such materials. Moreover, our method, accompanied by a deeper understanding of the theoretical model, can be utilized to study the Wannier functions and the specific matrix elements in crystals. These types of measurements can be of interest for several applications in solid state physics, such as in modern theory of polarization [21], or analysis of chemical bonds.

We acknowledge Diamond Light Source for time on Beamline I16 under Proposals [MT20485]. This work was supported by the Israel Science Foundation (ISF) (IL), Grant No. 201/18. The research leading to this result has been supported by the project CALIPSOplus under Grant Agreement 730872 from the EU Framework Programme for Research and Innovation HORIZON 2020.

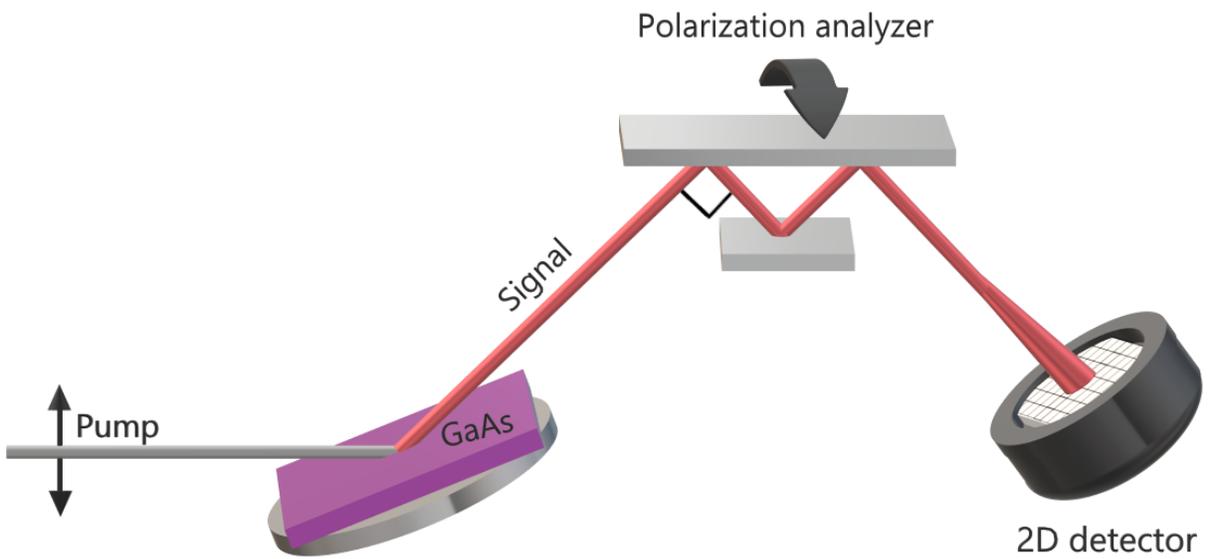

**Fig 1**: The experimental setup. The synchrotron radiation is polarized in the scattering plane and illuminates a GaAs crystal, which generates the PDC signal. This signal is separated from the background noise by a multi-bounce polarization analyzer and measured by a two-dimensional detector. The scattering plane of the analyzer can be rotated to filter different polarizations.

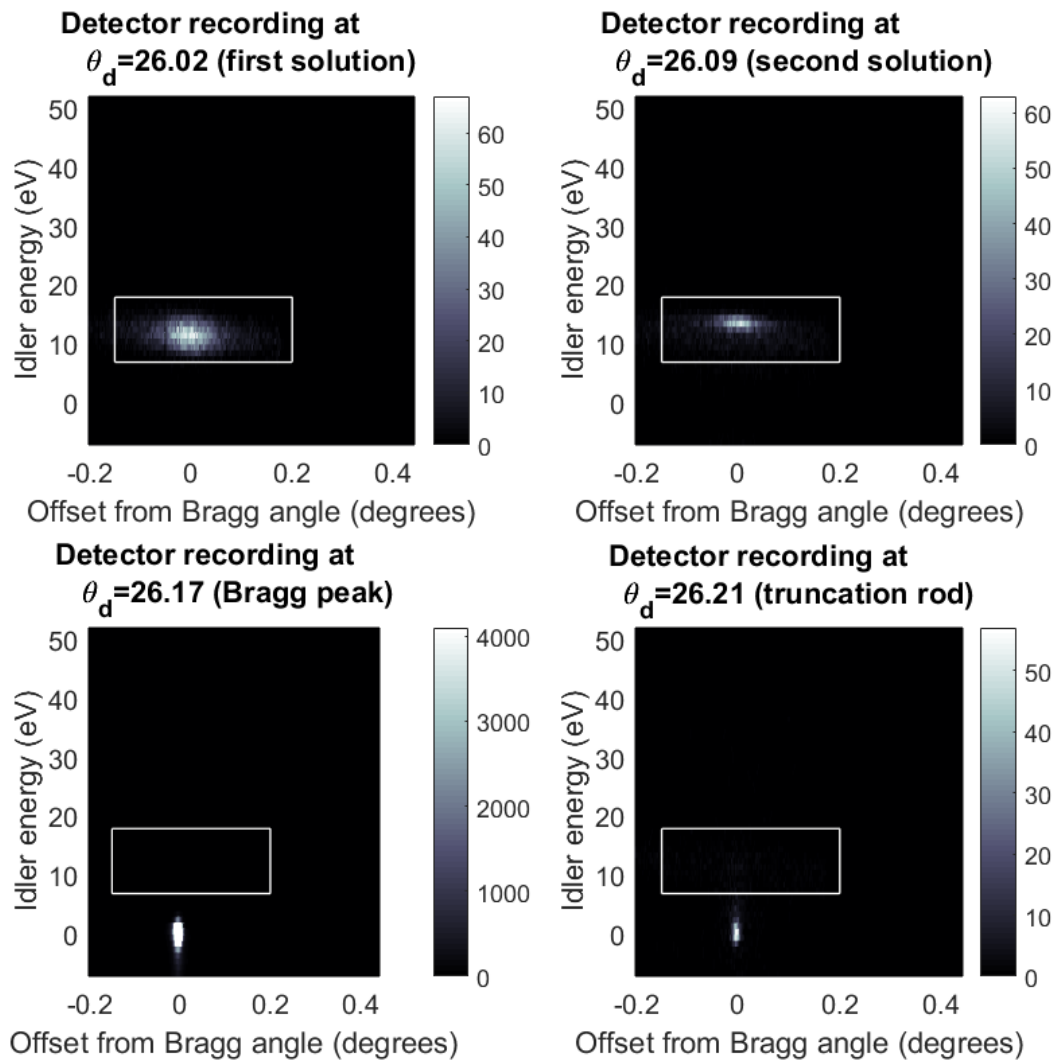

**Fig. 2**: Recorder raw data for (a) the peak of the first solution of the phase matching (detector angle at 26.06 degrees), (b) the peak of the second solution of the phase matching (detector angle at 26.09 degrees), (c) the Bragg peak (detector angle at 26.17 degrees), and (d) a truncation rod (detector angle at 26.21 degrees). The vertical direction on the detector is converted to energy deviation from the pump energy and the horizontal direction is converted to angle deviation from the Bragg angle. See further details in the text.

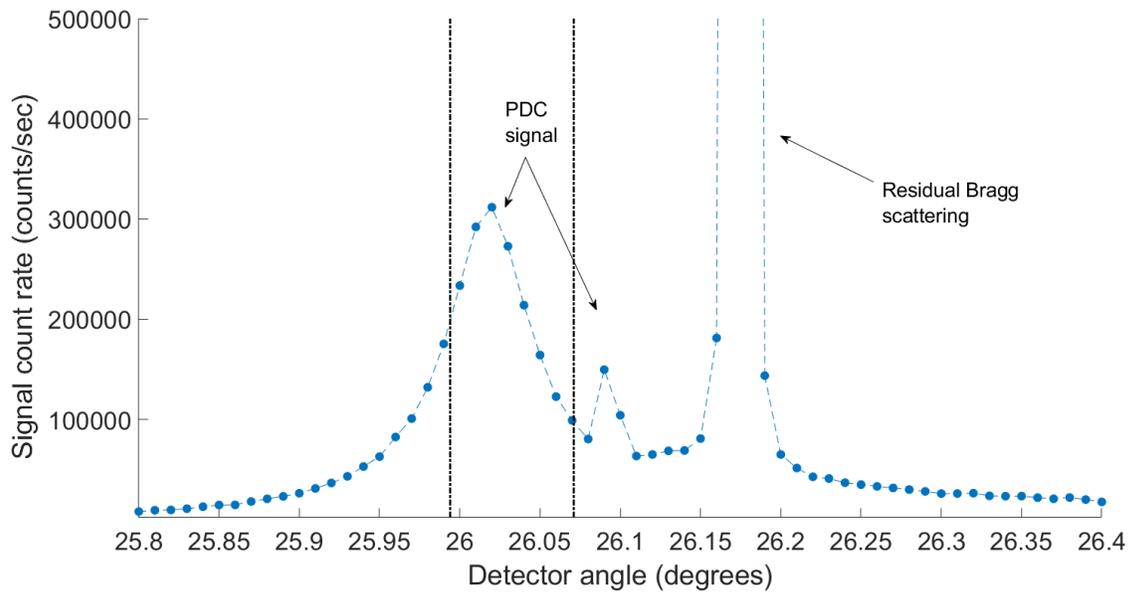

**Fig. 3**: X-ray signal count rate as a function of the detector angle. The idler energy is 10 eV. The two peaks on the left are the solutions for the phase matching equations and the strong peak on the right is the residual Bragg scattering. The dashed horizontal black line represents the calculated phase matching angles. The blue dashed line is a guide for the eye.

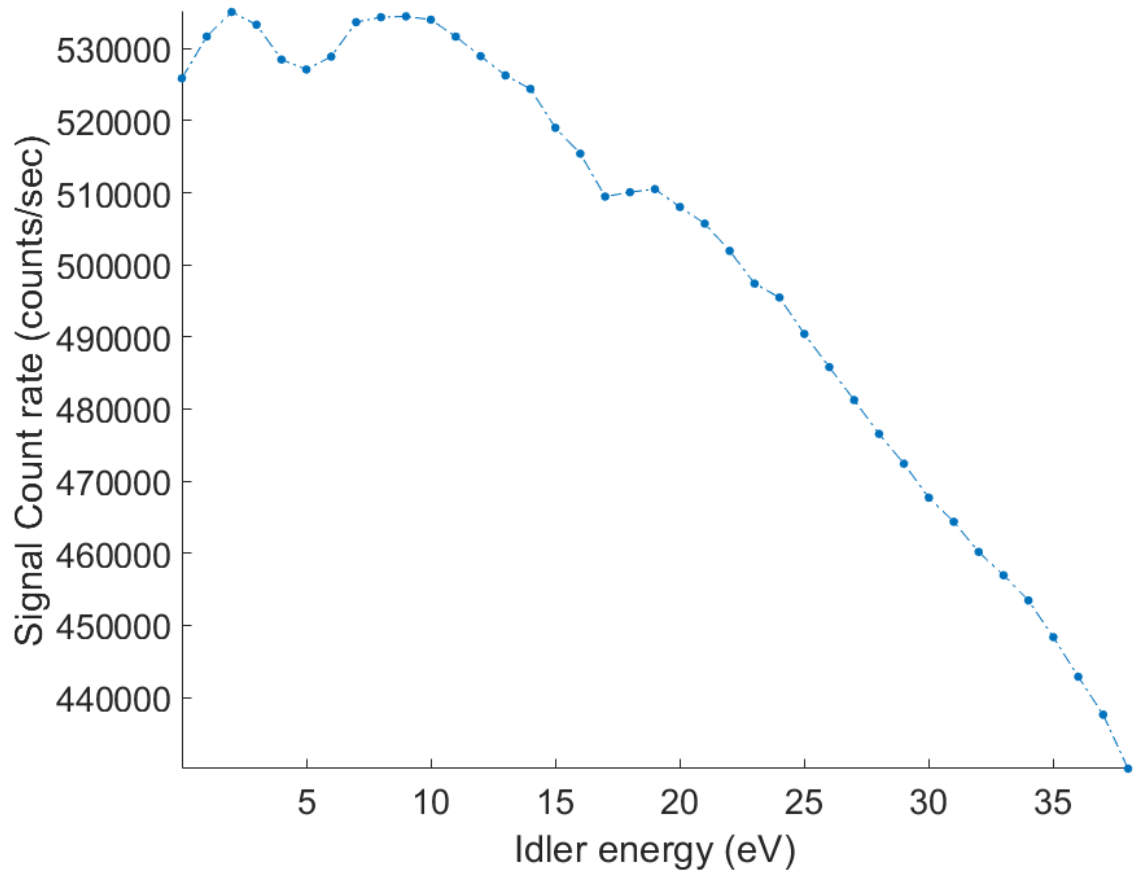

**Fig. 4**: Spectral measurement for the (1 1 1) atomic planes. The features around 1.4 eV and 6 eV can be attributed to the band gap energy and a transition energy within the band structure, respectively.

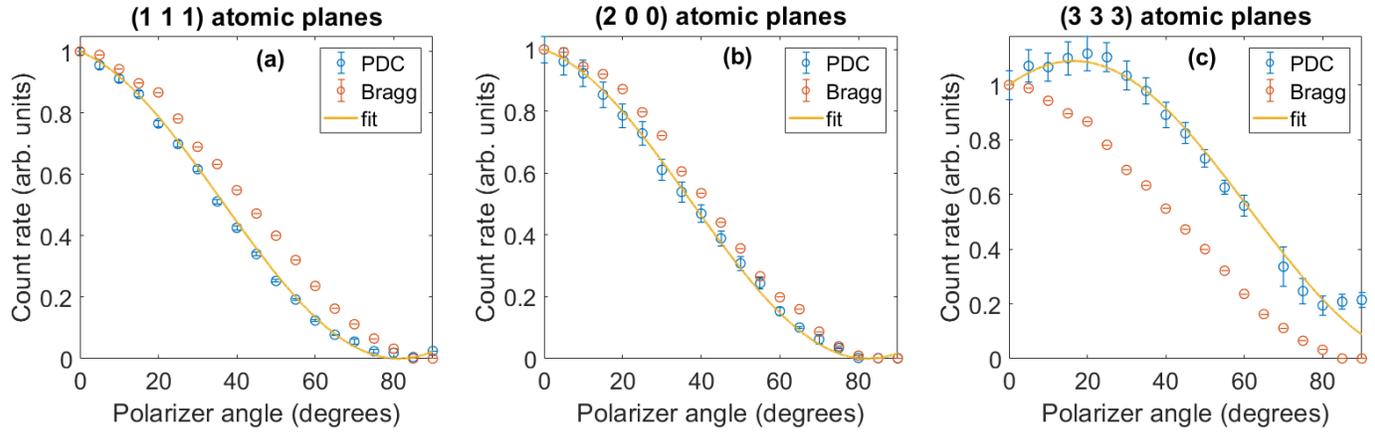

**Fig. 5**: Count rates as a function of the polarizer angle for the PDC signal photons with an idler energy at 10 eV for the (a) (1 1 1) atomic planes, (b) (2 0 0) atomic planes, and (c) (3 3 3) atomic planes of the GaAs sample. The blue dots are the measured PDC signal count rates, and the red dots are the measured count for the Bragg (elastic scattering). The solid line is a fit of a shifted squared cosine function for the count rate of the signal photons. The error bars are estimated by assuming Poissonian statistics.